\documentclass[]{pasj02} 
\usepackage[switch,mathlines]{lineno} 

\jyear{2025}
\Received{}
\Accepted{}

\usepackage{comment}

\begin{document} 

\title{A Bibliometric Analysis of the Scholarly Impact of Early Subaru Telescope-based Publications}

\author{Hideaki \textsc{Fujiwara}\orcid{0000-0001-6536-8656} }
\affil{Frontier Research Institute for Interdisciplinary Sciences (FRIS), Tohoku University, 6-3 Aramaki-aza Aoba, Aoba-ku, Sendai, Miyagi 980-8578, Japan}
\email{hideaki@fris.tohoku.ac.jp}


\maketitle

\begin{abstract}
Bibliometric methods provide valuable tools for assessing scientific productivity and impact across disciplines, yet their application in astronomy journals remains relatively limited. This study conducts a bibliometric analysis of Japanese astronomy publications before and after the commissioning of the Subaru Telescope, a major national investment in observational infrastructure. Using data from Scopus and SciVal, we examine peer-reviewed journal articles published between 1996 and 2007 by authors affiliated with Japanese institutions, focusing on field-normalized citation indicators such as the Field-Weighted Citation Impact (FWCI) and the share of publications in the top 10\% most cited globally. Subaru Telescope-based publications are identified through cross-referencing with official telescope publication lists and are compared against national and global benchmarks.
The results show that Subaru Telescope-based publications, while accounting for less than 10\% of Japan's total scholarly output in astronomy, consistently achieved FWCI values above 2.0 and a significantly higher proportion of highly cited papers. This indicates that the Subaru Telescope substantially enhanced Japan’s research visibility and impact, especially during its early operational years. This study demonstrates the utility of bibliometric evaluation in capturing the academic return of large-scale research facilities and contributes to broader discussions on research infrastructure in astronomy.
\end{abstract}


\section{Introduction}

Over the past several decades, bibliometric methods have become increasingly important for assessing research activity, collaboration patterns, and scholarly impact across a wide range of scientific disciplines. In astronomy, where large international facilities and collaborative projects are the norm, bibliometric indicators offer a quantitative complement to traditional qualitative assessments of research performance. Indicators such as publication volume, citation counts, and the proportion of highly cited papers are now commonly used to understand the influence of research outputs across countries, institutions, and subfields.

Japan has made significant contributions to observational astronomy through both national initiatives and international partnerships. Among these, the Subaru Telescope atop Maunakea, Hawaii, operated by the National Astronomical Observatory of Japan (NAOJ), stands out as a major investment in open-access, large-aperture ground-based astronomy. Since achieving scientific first light in 1999, the Subaru Telescope has enabled numerous discoveries and served as a platform for international collaboration and advanced instrumentation development. 

Bibliometric studies have previously been used to quantify the scientific return of ground-based large optical and infrared facilities, as well as the Hubble Space Telescope \citep{Benn2001,Trimble2005,Apai2010}. Although the scientific contributions of the Subaru Telescope have been qualitatively discussed by \citet{Oguri2025}, a bibliometric assessment specifically focused on its impact, particularly in relation to Japanese astronomy, remains underdeveloped. This study addresses that gap by quantitatively evaluating Subaru Telescope-based publications using indicators such as the Field-Weighted Citation Impact (FWCI) and the proportion of outputs in the top 10\% citation percentiles. It aims to clarify the broader role that major observatories play in shaping Japan’s research visibility.

While bibliometric analysis has been widely adopted in science policy, research evaluation, and other fields, its application in mainstream astronomy journals such as the Publications of the Astronomical Society of Japan (PASJ) has remained relatively limited. Nevertheless, bibliometric indicators can yield valuable insights when applied to specific astronomical contexts, particularly for evaluating the scientific return on large-scale facilities. The Subaru Telescope, as a high-profile international observatory led by Japan, offers a compelling case study for demonstrating the utility of such methods.

By analyzing scholarly outputs from Japanese-affiliated authors between 1996 and 2007, this study uses field-normalized citation indicators and collaboration metrics to assess the academic contribution of the Subaru Telescope in measurable terms. This also provides a broader perspective on the role of national observatories in enhancing international research visibility and fostering high-impact science. As astronomy increasingly depends on shared facilities, bibliometric methods will become more relevant in tracking long-term scientific trends and institutional performance.

We conduct a bibliometric analysis of peer-reviewed journal articles in astronomy and astrophysics authored by researchers affiliated with Japanese institutions between 1996 and 2007. Drawing on Scopus\footnote{https://www.scopus.com/} and SciVal\footnote{http://scival.com/} data, we examine trends in scholarly output and citation impact before and after the commissioning of the Subaru Telescope. In parallel, we identify and analyze Subaru Telescope-based publications, evaluating their citation performance.The findings offer insight into how large-scale observatories contribute to scholarly output, citation performance, and research visibility.

\section{Data and Methodology}\label{sec:2}

\subsection{Publication Database}\label{ssec:21}

Several major bibliographic databases are commonly used for literature analysis in astronomy and astrophysics, including the NASA Astrophysics Data System (ADS)\footnote{https://ui.adsabs.harvard.edu}, Web of Science (WoS)\footnote{https://webofknowledge.com/wos/}, and Scopus. ADS is widely regarded within the astronomical community due to its extensive coverage of astrophysical journals and integration with data archives. Web of Science and Scopus, on the other hand, are multidisciplinary databases that provide standardized citation metrics and support large-scale bibliometric analyses.

In this study, we used Scopus as our primary data source, owing to its broad coverage of peer-reviewed journals, standardized subject classification (ASJC codes), and compatibility with SciVal, a dedicated analytics platform developed by Elsevier. SciVal allows for the retrieval of field-normalized citation metrics such as FWCI, and supports structured comparisons across institutions, countries, and time periods. These features make Scopus and SciVal particularly suitable for the longitudinal and comparative analysis conducted in this work.

\subsection{Dataset Construction}\label{ssec:22}

The dataset for this study was derived from SciVal, based on Scopus content, by filtering for publications classified under the subject area ``Astronomy and Astrophysics'' (ASJC code 3103). Only peer-reviewed journal articles (publication type: ``Article'') were included; other publication types such as reviews, conference proceedings, editorials, and short communications were excluded. This approach ensures a consistent focus on original research outputs with comparable citation behavior.

The target publication period spans from 1996, the earliest year for which field-normalized citation indicators (such as FWCI) are available in SciVal, through 2007. This end point allows for a medium-term citation window while capturing the early operational years of the Subaru Telescope. To facilitate comparative analysis, the data were grouped into three four-year intervals:  
(1) 1996–1999, representing the pre-Subaru Telescope period;  
(2) 2000–2003, corresponding to the initial operational phase of the Subaru Telescope; and  
(3) 2004–2007, representing a phase of stable operations of the Subaru Telescope.

Conference papers--including those published in the SPIE Conference Proceedings, a key outlet for astronomical instrumentation--were excluded due to their distinct citation patterns and the challenges they pose for bibliometric normalization. While instrumentation research is central to observational astronomy, its evaluation requires a different analytical framework and is reserved for future work.

Both Scopus and SciVal data are updated on a regular basis; however, for consistency and reproducibility, all analyses in this study were based on the dataset available as of the last update on June 11, 2025.

\subsection{Identification of Subaru Telescope-based Publications}\label{ssec:23}

To identify published papers in which data obtained by the Subaru Telescope were used (whether primarily or partially), we utilized the official annual publication lists\footnote{https://subarutelescope.org/Science/Publish/} maintained by the Subaru Telescope. These lists provide bibliographic records of peer-reviewed articles that explicitly acknowledge the use of Subaru Telescope data. The earliest such papers, notably including the milestone paper by \citet{Kaifu2000}, were published in February 2000, following the telescope’s engineering first light in December 1998 and scientific first light in January 1999.

Using web scraping techniques, we extracted metadata such as article title, authorship, journal reference, and ADS links from these lists. Where available, Digital Object Identifiers (DOIs) were retrieved via the NASA ADS API using bibcodes embedded in the provided URLs. The resulting master list of Subaru Telescope-based publications was then cross-matched with our Scopus dataset using DOIs and/or article titles.
In cases where DOIs were not assigned, we manually located the corresponding entries in Scopus and retrieved their Electronic IDs (EIDs) for integration. Articles for which only arXiv preprints were available without corresponding journal publications were excluded. Similarly, publications in journals not indexed by Scopus were omitted. These steps ensured that all Subaru Telescope-based publications included in our analysis were bibliometrically traceable within the same framework as the rest of the dataset.

It should be noted that while the official Subaru Telescope publication lists are maintained with care and are considered reliable, they are not intended to be exhaustive. Therefore, the resulting dataset may not include all Subaru Telescope-based publications, particularly those that did not explicitly acknowledge data usage or were omitted from the listing for other reasons.

\subsection{Entity Definition and Comparison Sets}\label{ssec:24}

\begin{table*}
  \tbl{Number of publications in each publication set defied in this study.}{%
  \begin{tabular}{lccc}
      \hline
      Publication Set & 1996--1999 & 2000--2003 & 2004-2007  \\ 
      \hline
      ``Astronomy and Astrophysics'' publications & $38252$ & $38817$ & $43860$ \\
      ``Astronomy and Astrophysics'' publications from Japan & $2748$ & $3073$ & $3116$ \\
      Subaru Telescope-based publications from Japan & -- & $113$ & $246$ \\
      \hline
    \end{tabular}}\label{tab:first}
\begin{tabnote}
\end{tabnote}
\end{table*}

For the purpose of this study, we focused specifically on the scholarly contribution of the Japanese astronomy community. To this end, we filtered the Subaru Telescope-based publication set to include only those papers with at least one author affiliated with a Japanese institution. This filtered subset---referred to as ``Subaru Telescope-based publications from Japan''---serves as the core analytical entity in all subsequent analyses.

In SciVal, this entity was created by applying a ``Countries and Regions'' filter to the broader Subaru-based publication set, restricting it to authors affiliated with Japanese institutions. This allowed for standardized benchmarking using SciVal's metrics and ensured consistency with national-level comparisons.

For comparative purposes, we also defined two reference sets:  
(1) All journal articles categorized under the subject area ``Astronomy and Astrophysics'' (ASJC code 3103), representing global scholarly output in this field; and  
(2) A filtered subset of the same category limited to publications with at least one author affiliated with a Japanese institution.

These comparison sets allow us to assess the relative scholarly performance of Subaru Telescope-based publications within both national and international contexts. Table~\ref{tab:first} summarizes the number of identified publications in each of the three publication sets defined in this study. The sets are nested such that the Subaru Telescope-based publications from Japan form a subset of Japanese-authored publications, which in turn are a subset of all global publications in the field.

\subsection{Bibliometric Indicators}\label{ssec:25}

\begin{table*}
  \caption{Bibliometric indicators used in this study.}
  \label{tab:indicators}
  \centering
  \begin{tabular}{lp{9cm}}
    \hline
    \textbf{Indicator} & \textbf{Short Description} \\
    \hline
    Scholarly Output & The number of publications of a selected entity. \\
    Citation Count & Total citations received by publications of the selected entities. \\
    Field-Weighted Citation Impact (FWCI) & The ratio of citations received relative to the expected world average for the subject field, publication type and publication year. World FWCI is 1.0. \\
    Outputs in Top 10\% Citation Percentiles & The number of publications of a selected entity that are highly cited, having reached a threshold of citations received.  \\
    International Collaboration & The extent of international co-authorship. \\
    \hline
  \end{tabular}
\end{table*}

To assess temporal trends in publication activity and scholarly impact, we extracted annual bibliometric indicators from SciVal for the period 1999--2007. 
The selected indicators are: 
(1) the number of publications (Scholarly Output); 
(2) total citations received (Citation Count); 
(3) Field-Weighted Citation Impact (FWCI); 
(4) the proportion of publications within the top 10\% globally by FWCI (Outputs in Top 10\% Citation Percentiles, field-weighted); and 
(5) the extent of international co-authorship (International Collaboration). 
Each of these indicators provides a distinct perspective on research performance and is widely used in bibliometric studies to evaluate scientific productivity, visibility, and influence. Table~\ref{tab:indicators} summarizes the definitions and analytical roles of these indicators. 

Annual values of the indicators were extracted from SciVal for the following three entities, restricted to peer-reviewed journal articles (publication type: ``Article''): 
(1) publications categorized under ``Astronomy and Astrophysics'' (ASJC code 3103); 
(2) publications in the same category with at least one author affiliated with a Japanese institution; and 
(3) Subaru Telescope-based publications from Japan. 
In addition, Citation Count and FWCI were also retrieved at the individual publication level for detailed analysis.

\section{Results}\label{sec:3}
\subsection{Trends in Scholarly Output}\label{ssec:31}

\begin{figure*}
 \begin{center}
  \includegraphics[width=125mm]{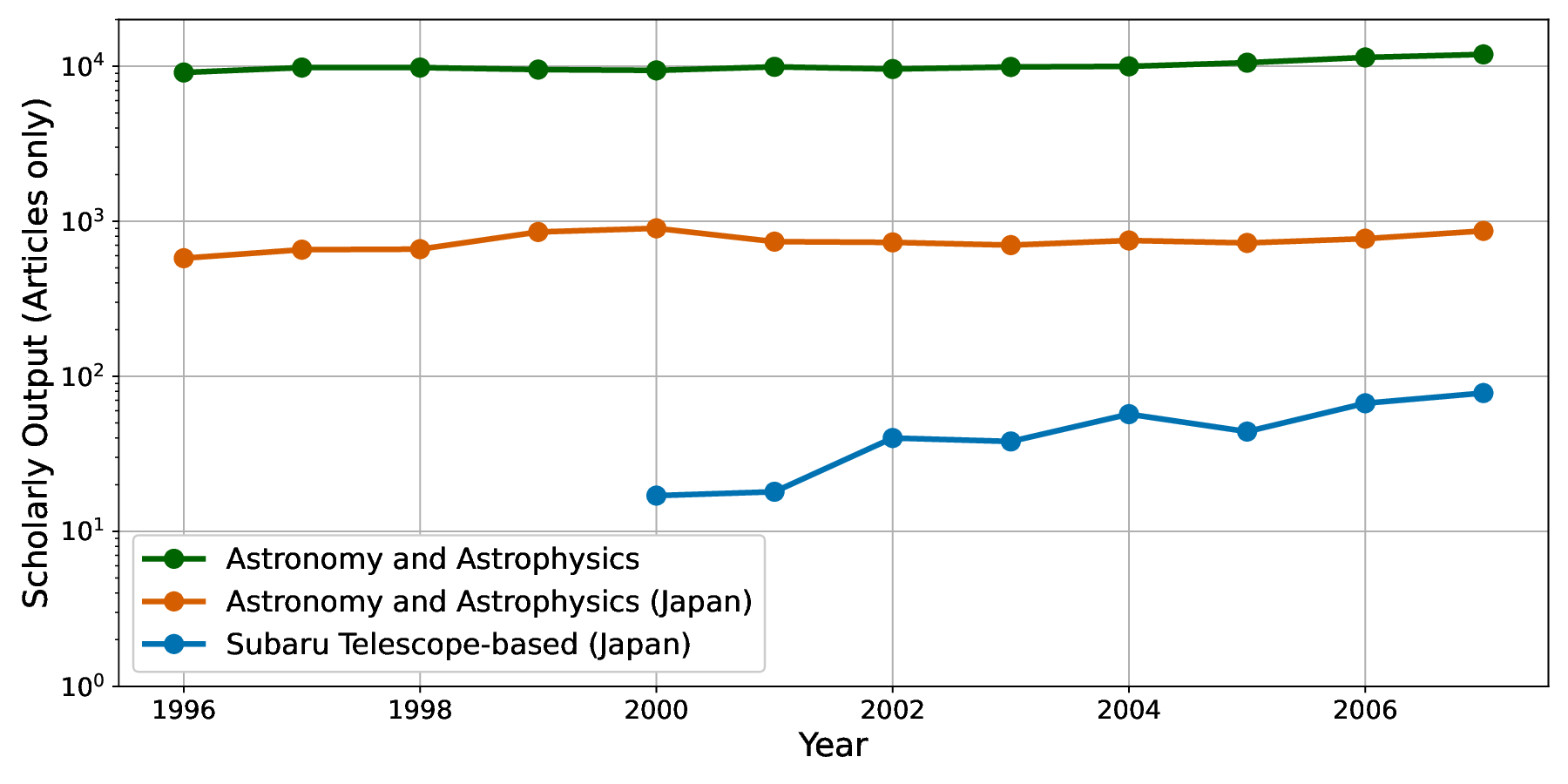}
 \end{center}
 \caption{
Annual number of peer-reviewed journal articles (Scholarly Output) from 1996 to 2007 across three publication sets: global Astronomy and Astrophysics publications (green), publications from Japan (red), and Subaru Telescope-based publications from Japan (blue). The vertical axis uses a logarithmic scale. While global Scholarly Output increased steadily and Japan’s remained relatively stable, Subaru Telescope-based publications rose rapidly after 2000. Despite accounting for less than 10\% of Japan’s total output, this growth reflects the telescope’s increasing scientific use during its early operational years and highlights its expanding role in Japan’s astronomy research.
{Alt text: A line graph with a logarithmic vertical axis. The x-axis shows publication years from 1996 to 2007. The y-axis represents the number of peer-reviewed journal articles, on a logarithmic scale from $10^0$ to approximately $10^4$. Three lines represent global publications, Japanese publications, and Subaru Telescope-based publications from Japan.}
}
  \label{fig:scholarly_output}
\end{figure*}

Figure~\ref{fig:scholarly_output} shows the annual number of peer-reviewed journal articles (Scholarly Output) from 1996 to 2007 for three publication sets: global publications in “Astronomy and Astrophysics,” publications from Japan in the same field, and Subaru Telescope-based publications from Japan. Over this period, the global Scholarly Output increased steadily from 9117 articles in 1996 to 11927 in 2007. In contrast, Japan’s Scholarly Output remained relatively stable, ranging from approximately 700 to 900 articles per year. Subaru Telescope-based publications from Japan began appearing in 2000 and increased from 17 articles in that year to 78 in 2007. While Subaru Telescope-based publications accounted for less than 10\% of Japan’s annual Scholarly Output in astronomy throughout the period, their rapid increase is noteworthy. This trend highlights the expanding scientific use of the Subaru Telescope during its early operational years and its growing role within Japan’s astronomy research landscape.

\subsection{Citation Impact}\label{ssec:32}

Citation impact was assessed using two complementary field-normalized metrics: Field-Weighted Citation Impact (FWCI) and Outputs in Top 10\% Citation Percentiles. These indicators provide distinct yet complementary perspectives on scholarly influence. FWCI measures the average citation performance of a publication set relative to global expectations for similar publications, while the Top 10\% indicator quantifies the proportion of papers that rank among the top 10\% most cited publications in their field, year, and publication type.

\begin{figure*}
 \begin{center}
  \includegraphics[width=125mm]{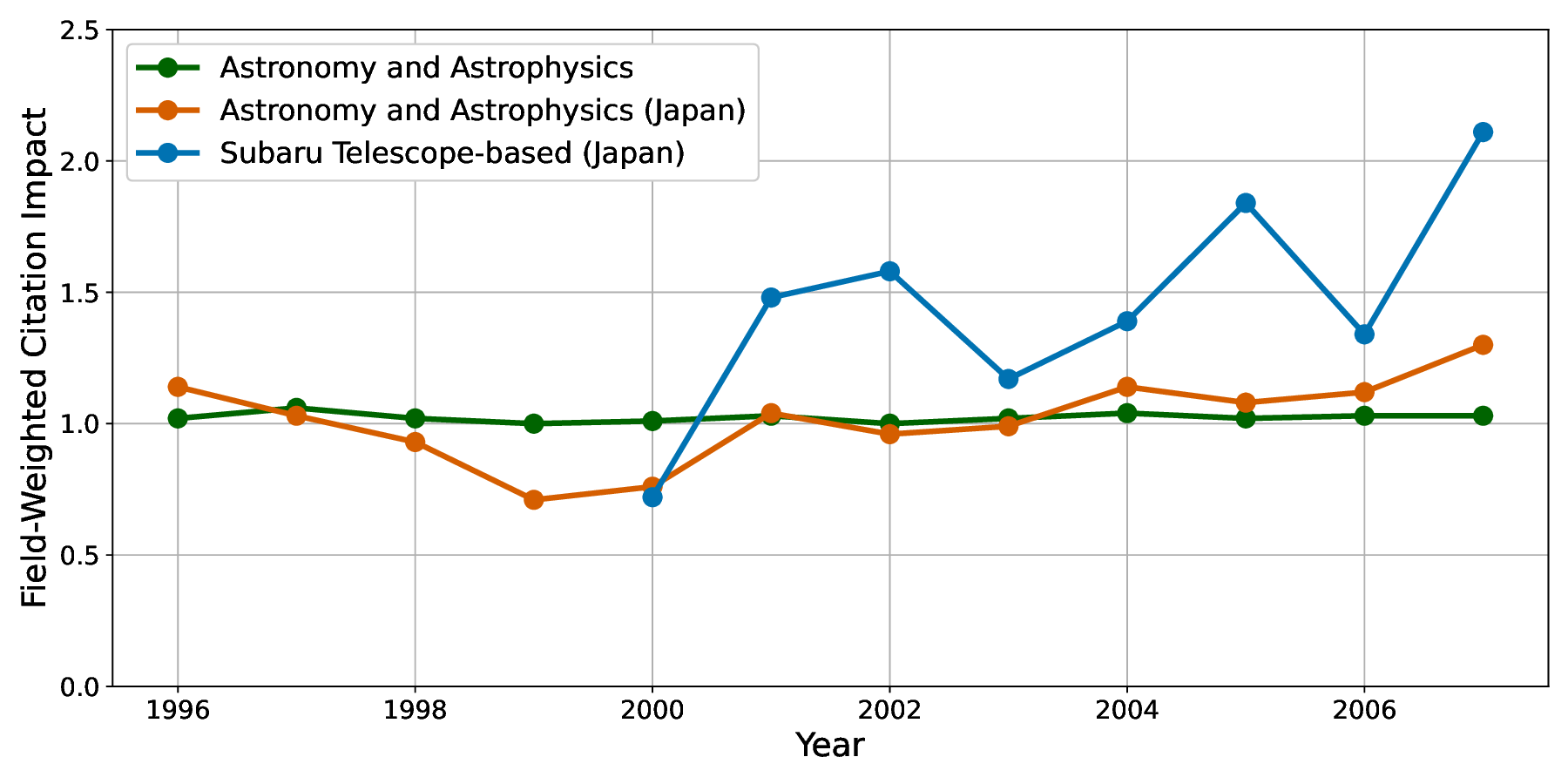}
  \includegraphics[width=124mm]{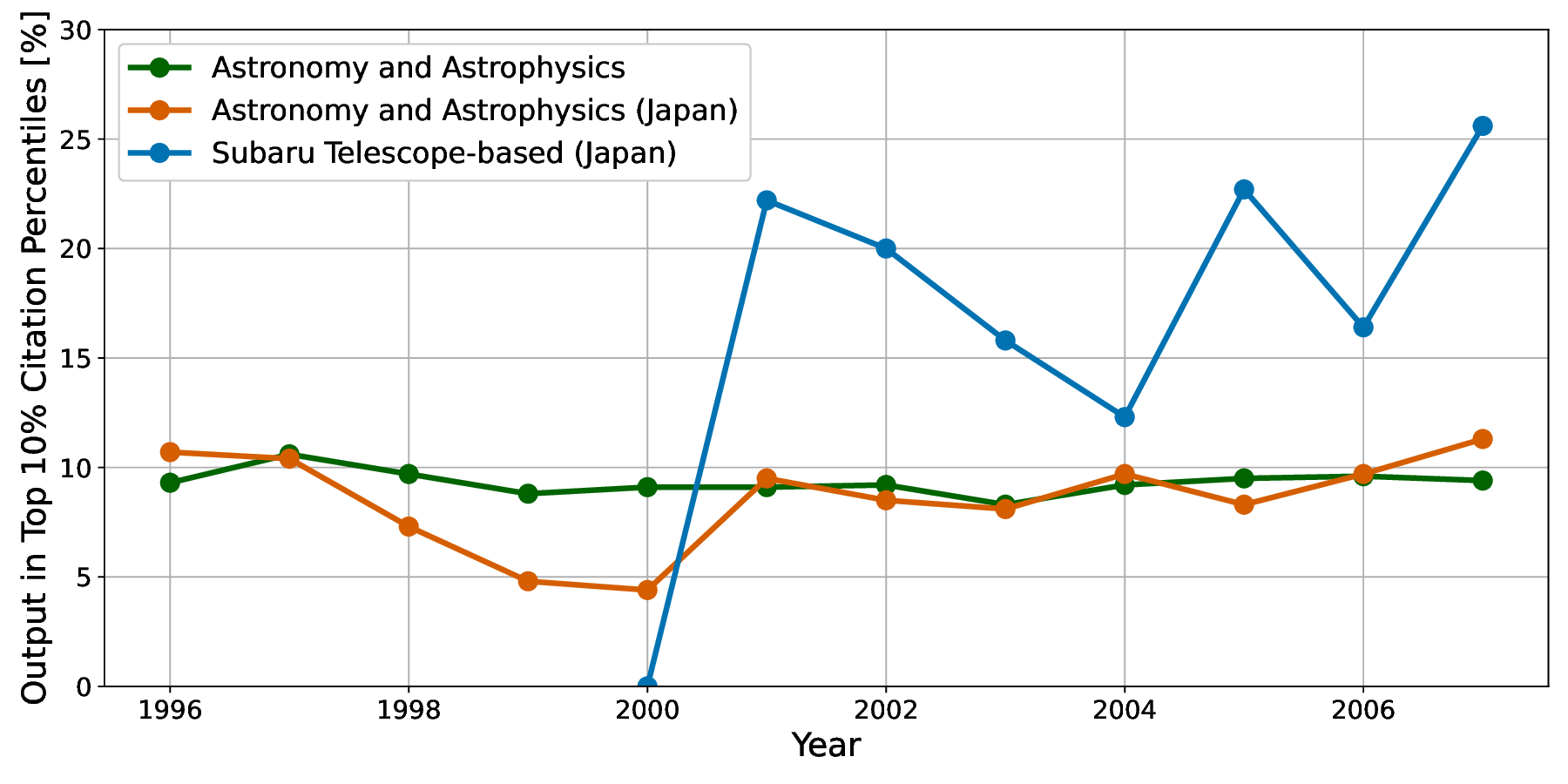}
 \end{center}
\caption{
Annual trends in citation impact for peer-reviewed journal articles in astronomy and astrophysics from 1996 to 2007. 
(Top) Field-Weighted Citation Impact (FWCI). A value of 1.0 represents the global average, normalized by field, publication year, and publication type. 
(Bottom) Proportion of publications in the Top 10\% Citation Percentiles. The global benchmark remains near 10\% by definition.
In both panels, blue denotes Subaru Telescope-based publications from Japan, red represents all Japan-affiliated publications, and green shows the global baseline. 
Subaru Telescope-based publications consistently outperformed both national and global averages in both indicators, indicating strong relative citation performance and a higher frequency of high-impact research.
{Alt text: Two line graphs with linear vertical axes. In both panels, the x-axis shows publication years from 1996 to 2007. The top panel displays Field-Weighted Citation Impact (FWCI), ranging from 0.0 to 2.5. The bottom panel shows the proportion of publications in the Top 10\% Citation Percentiles, ranging from 0\% to 30\%. Each panel includes three lines representing global publications, Japanese publications, and Subaru Telescope-based publications from Japan.}
}
  \label{fig:fwci-and-top10}
\end{figure*}

Figure~\ref{fig:fwci-and-top10} (top panel) shows the annual trends in FWCI for the three publication sets. As expected, the global FWCI remained close to 1.0 throughout the study period. Japan's FWCI values were generally lower than the global average, especially between 2000 and 2004, suggesting a relatively modest citation performance. In contrast, Subaru Telescope-based publications from Japan consistently recorded higher FWCI values, exceeding 2.0 in 2007. This indicates that these publications received, on average, more than twice the expected number of citations for their type and context.

A similar trend is evident in the Top 10\% indicator, as illustrated in Figure~\ref{fig:fwci-and-top10} (bottom panel). The global proportion of Outputs in Top 10\% Citation Percentiles was stable at approximately 10\% by definition. The corresponding proportion for Japanese publications was consistently lower, with only 4.4\% in 2000, for example. However, Subaru Telescope-based publications from Japan consistently exhibited much higher rates—22.2\% in 2001 and 25.6\% in 2007—demonstrating a significantly greater concentration of highly cited papers.

Taken together, these indicators demonstrate that Subaru Telescope-based publications from Japan achieved substantially higher scholarly impact compared to both national and global averages. Although these publications accounted for less than 10\% of Japan’s annual Scholarly Output in astronomy during this period, they made a disproportionately large contribution to high-impact research.

\subsection{International Collaboration}\label{ssec:33}

In bibliometric studies, international collaboration, typically measured as the proportion of publications co-authored by researchers affiliated with institutions in different countries, is widely recognized as an indicator of research connectivity and visibility. Prior research has shown that such collaborations are often associated with higher citation rates, likely due to broader dissemination, diverse expertise, and expanded institutional networks \citep{Glanzel2001,Gazni2012}. Accordingly, examining international collaboration provides valuable context for interpreting citation-based impact metrics. Although a detailed analysis of collaboration networks is beyond the scope of this study, we briefly examined the proportion of internationally co-authored journal articles across the three publication sets. The results show a steady increase in international collaboration from 1996 to 2007, both globally and in Japan. Specifically, the global share of internationally co-authored publications in astronomy and astrophysics rose from approximately 34\% in 1996 to 48\% in 2007, while Japan’s share increased from about 45\% to 62\%.

Subaru Telescope-based publications from Japan showed particularly high levels of international co-authorship throughout the study period. In 2000, 94.1\% of these publications were classified as international collaborations, and from 2001 onward the rate consistently exceeded 80\%, with representative values of 86.8\% in 2003, 90.9\% in 2005, and 89.7\% in 2007. However, caution is warranted in interpreting these figures. Researchers affiliated with the Subaru Telescope in Hawaii, which is operated by NAOJ, often list U.S.-based institutional addresses in publications, even when they are Japanese nationals and full-time NAOJ staff. As a result, bibliometric databases such as Scopus and SciVal may classify these publications as internationally co-authored, even when all contributors are institutionally domestic. This classification artifact was further amplified by the operational structure of Subaru Telescope in its early years, when including Hawaii-based engineers and scientists as co-authors was common and often essential due to their key roles in instrumentation, observations, and operations. While the consistently high international co-authorship rates reflect the global role of the Subaru Telescope, part of this trend likely results from affiliation-based classification effects rather than purely cross-national collaboration.

\section{Discussion}\label{sec:4}

\subsection{Contextualizing Citation Impact}\label{ssec:41}

The findings presented above indicate a marked shift in the citation impact of Japanese astronomy publications following the commissioning of the Subaru Telescope. While Japan’s overall Scholarly Output in astronomy remained relatively constant between 1996 and 2007, the emergence and growth of Subaru Telescope-based publications were associated with substantially higher field-normalized citation performance.

Elevated FWCI values and a consistently high share of Outputs in Top 10\% Citation Percentiles suggest that publications utilizing Subaru data attained greater visibility and scholarly influence than both the national and global baselines. This pattern points to the Subaru Telescope’s role not only in enabling new research capacity, but also in enhancing the international standing of Japanese astronomy.

Several factors likely contributed to this increased impact. The Subaru Telescope offered unique technical capabilities, such as wide-field optical imaging with Suprime-Cam \citep{Miyazaki2002} and high-resolution spectroscopy with the High Dispersion Spectrograph (HDS) \citep{Noguchi2002}, that enabled competitive science and attracted ambitious research projects. Its ``open-sky'' policy and emphasis on international collaboration further facilitated global partnerships, as evidenced by the high international co-authorship rates.

Beyond its immediate bibliometric footprint, the telescope may have supported longer-term capacity building by expanding opportunities for domestic researchers to engage in frontier science and establish international networks. The rise in impact also coincided with a broader shift in astronomy toward globally integrated, facility-driven research, exemplified by the increasing role of large-scale observatories and collaborative infrastructures.

These results underscore the wider role of large-scale research infrastructures not only as enablers of discovery but also as instruments for strengthening national research visibility and competitiveness. The case of the Subaru Telescope demonstrates how a focused subset of publications linked to a single facility can exert a disproportionate influence on national-level scholarly performance metrics.

\subsection{Topical Shifts in Scholarly Output}\label{ssec:42}

\begin{figure*}
 \begin{center}
  \includegraphics[width=180mm]{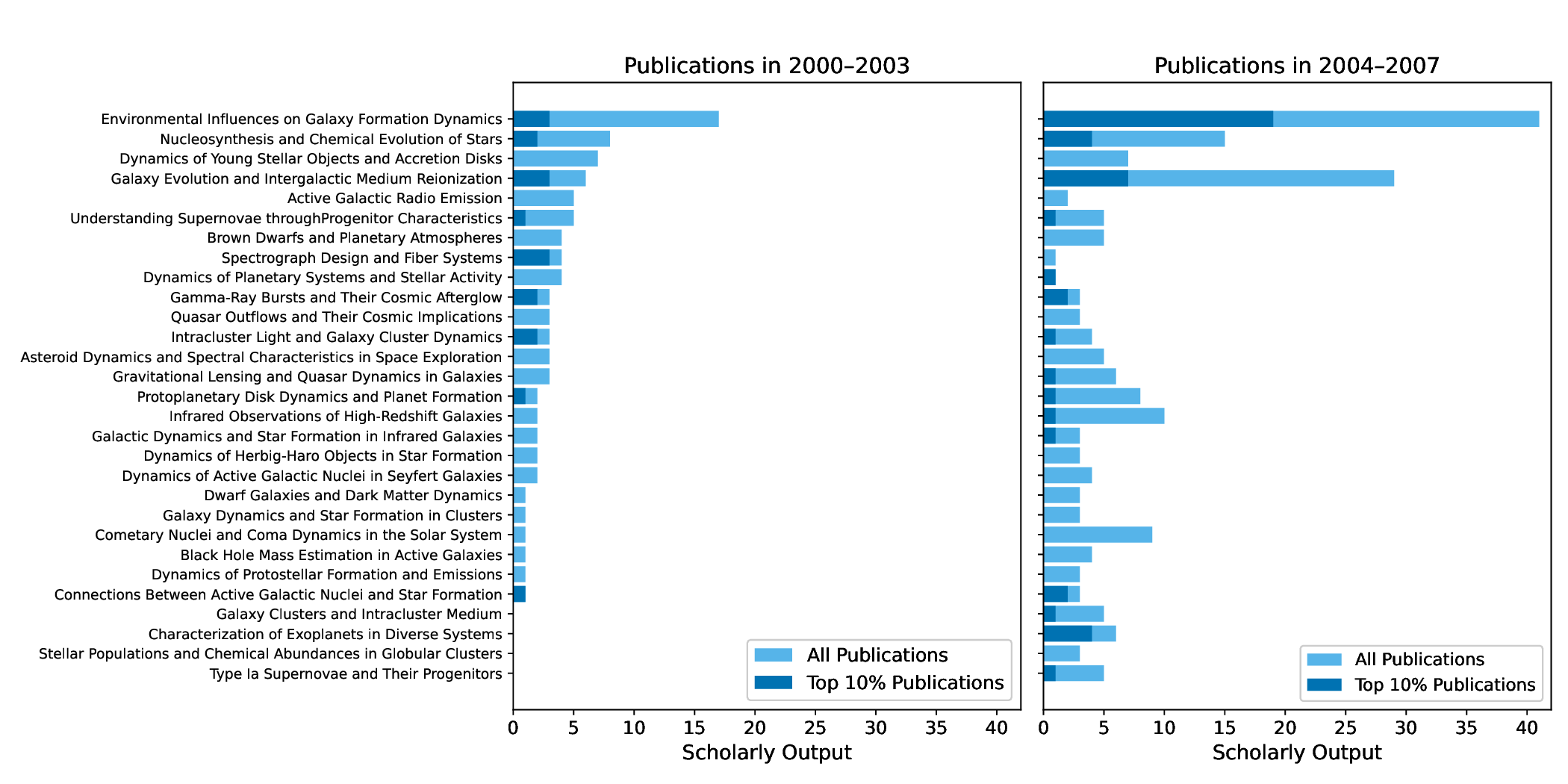}
 \end{center}
   \caption{  Topic-wise distribution of Subaru Telescope-based publications from Japanese institutions during two periods: 2000--2003 (left) and 2004--2007 (right). For each topic, the total number of publications (light blue) and the number that rank among the top 10\% most cited worldwide (dark blue) are shown. Only topics with at least three publications in either period are included. Topics are sorted by total output in 2000--2003. The comparison highlights shifts in thematic focus and relative citation impact over time.
{Alt text: Two grouped bar charts comparing topic-wise distributions of Subaru Telescope-based publications from Japan. The left chart covers the period 2000–2003 and the right chart covers 2004--2007. Each topic has two bars: one for total publications and one for publications in the Top 10\% Citation Percentiles. Topics are ordered by publication volume in 2000--2003.}
}
  \label{fig:topic}
\end{figure*}

A topical breakdown of Subaru Telescope-based publications from Japan (Figure~\ref{fig:topic}) illustrates how the scientific focus of highly cited work evolved over time. In both periods (2000--2003 and 2004--2007), topics such as galaxy formation and stellar evolution remained prominent. However, new areas of interest such as high-redshift galaxies and planet formation gained visibility in the latter period. This shift likely reflects the broader range of scientific investigations made possible by the commissioning of new instruments that came online around 2000. The topic categories used here are automatically assigned by SciVal based on article-level clustering and are intended to provide a general thematic overview; their granularity and consistency may vary across fields.

This analysis underscores how large-scale facilities can shape the thematic evolution of research over time. Future studies may extend this approach by using longer citation windows, incorporating more recent data, or analyzing additional publication types such as instrumentation papers or conference proceedings. The bibliometric framework developed here could also be applied to other observatories to better evaluate their scientific return.

\subsection{Tracking Citation Performance of Subaru Telescope-based Publications}\label{ssec:43}

\begin{figure*}
 \begin{center}
  \includegraphics[width=85mm]{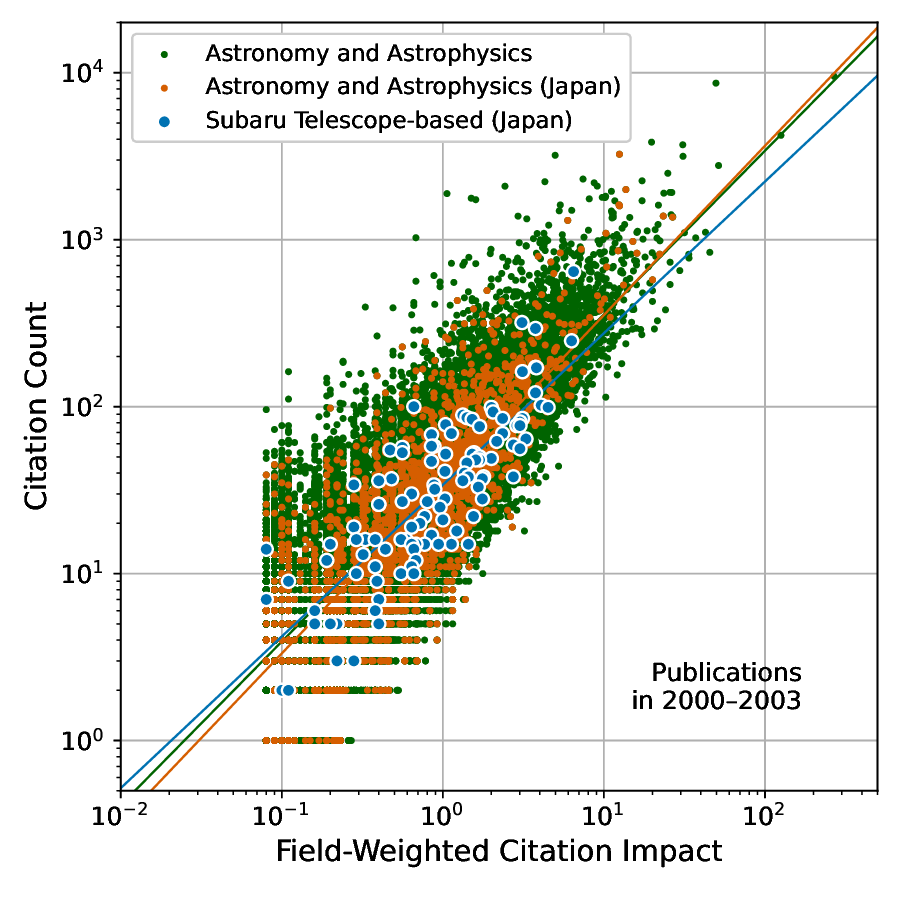}
  \includegraphics[width=85mm]{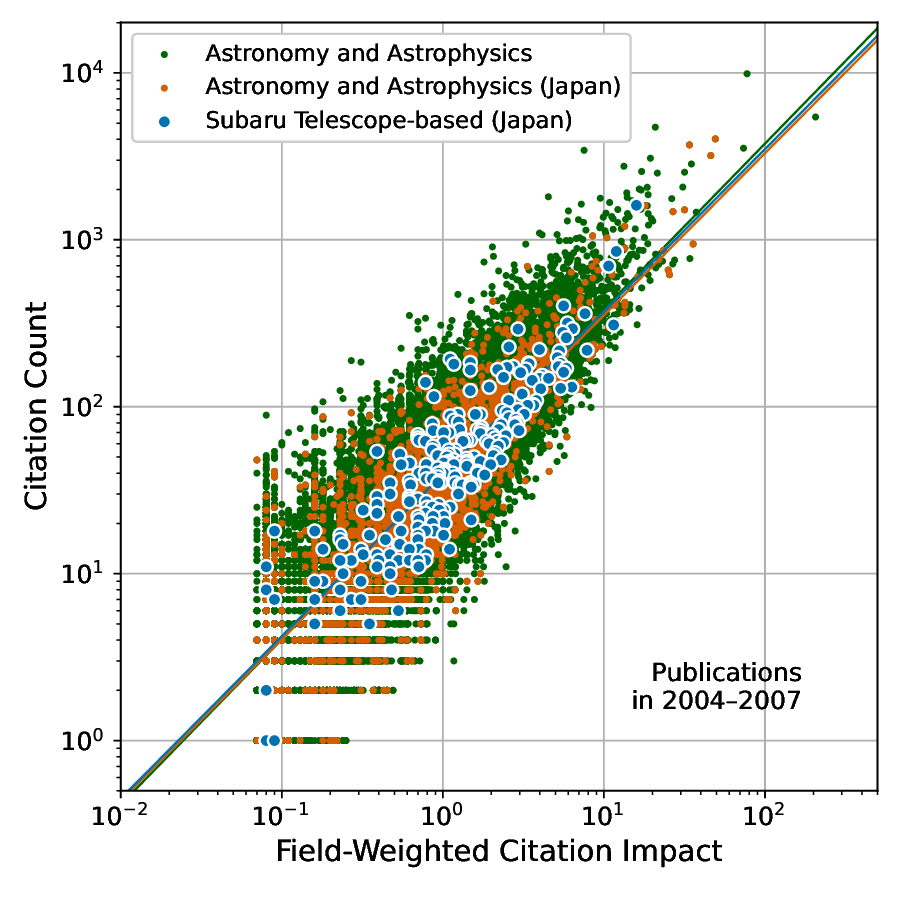}
 \end{center}
  \caption{
Total citation counts versus FWCI for individual publications from 2000--2003 (left) and 2004--2007 (right). Each point represents a peer-reviewed journal article from one of three publication sets: global astronomy and astrophysics literature (green), publications from Japanese institutions (red), and Subaru Telescope-based publications from Japan (blue, with larger, white-edged markers). The vertical axis indicates the total number of citations accrued up to the time of data extraction (June 2025), while FWCI is calculated based on citations received within a fixed three-year window following the publication year. A strong correlation is observed across all sets, although the scatter also reveals variation between early and long-term impact. This comparison highlights how early citation attention can vary depending on publication timing and context. Linear regression lines, fitted in log-log space and color-matched to each publication set, are overlaid to visualize the general trends.
{Alt text: Two scatter plots. The left panel shows data for 2000--2003, and the right panel for 2004--2007. In both panels, the x-axis shows FWCI ranging from approximately $10^{-2}$ to $5 \times 10^2$, and the y-axis shows total citation counts ranging approximately from $10^0$ to about $10^4$. Each point represents a single publication from one of three sets: global, Japan, and Subaru Telescope-based publications from Japan. Subaru  Telescope-publications are marked with larger, white-edged dots. Log-log linear regression lines are shown for each group.}
}
  \label{fig:fwci_vs_citations}
\end{figure*}

FWCI is widely used to normalize differences in citation behavior across disciplines, publication years, and publication types. Figure~\ref{fig:fwci_vs_citations} shows total citation counts versus FWCI for individual publications from the periods 2000--2003 and 2004--2007. In general, articles with higher total citation counts also exhibit higher FWCI values. To clarify this relationship, linear regression lines were fitted in log-log space for each publication set. This is expected, as FWCI is calculated from citations received within a fixed three-year window following publication, while total citation counts reflect all citations accumulated to date. Accordingly, FWCI can be viewed as a proxy for early citation performance, whereas total citation count represents long-term scholarly impact.

Despite this correlation, differences in slope and scatter between the two periods suggest temporal shifts in citation behavior. Notably, Subaru Telescope-based publications from 2000--2003 exhibit a shallower regression slope than other sets, indicating that these papers tended to receive relatively high FWCI scores even at moderate total citation levels. By contrast, Subaru Telescope-based papers from 2004--2007 display a slope more consistent with broader trends.

One possible interpretation of this pattern is that the earliest wave of Subaru Telescope-based publications, those released shortly after first light, garnered rapid citation attention, reflecting the community's initial interest in the new facility and its early scientific output. As the telescope became more established, citation behavior surrounding Subaru Telescope-based work appears to have normalized, aligning more closely with general trends in the field. This shift highlights how the novelty and anticipation surrounding a new research infrastructure can temporarily enhance early citation performance, producing elevated FWCI values even before long-term impact is fully realized.

\section{Conclusion}\label{sec:5}

This study presented a bibliometric evaluation of the scholarly impact of the Subaru Telescope on Japanese astronomy by analyzing peer-reviewed publications from 1996 to 2007. Using field-normalized citation indicators such as the Field-Weighted Citation Impact (FWCI) and the Top 10\% citation percentile, we found that Subaru Telescope-based publications consistently outperformed both national and global benchmarks in terms of citation impact and international collaboration.

While the overall number of Japanese astronomy publications remained relatively stable over the study period, Subaru Telescope-based papers exhibited significantly higher visibility and influence, particularly during the telescope’s early operational years. These findings underscore not only the scientific value of the Subaru Telescope, but also its role in enhancing Japan’s international research presence.

Field-normalized metrics proved to be an effective lens for assessing the comparative performance of facility-driven research. Our results highlight how large-scale astronomical infrastructure can strengthen national research capacity, foster high-impact outputs, and elevate global visibility.

Future work may extend this analysis by incorporating more recent publication data, particularly from the era following the commissioning of high-impact instruments such as Hyper Suprime-Cam (HSC) \citep{Aihara2018,Miyazaki2018}. Additionally, comparative bibliometric studies involving other large ground-based telescopes of similar scale could help to further contextualize the performance of the Subaru Telescope and identify distinctive patterns in its scientific output and citation impact.

\begin{ack}
The author thanks Drs. Kazuyuki Suzuki, Marc Hansen, and Michiaki Yumoto for their generous guidance on bibliometric methods, which greatly helped shape the foundation of this study. 
The author is also grateful to Dr.\ Toshiyuki Hayase for frequent discussions on bibliometric approaches. 
The author's experience as both a user and a former scientist at the Subaru Telescope provided the original motivation for this research, offering insights grounded in practical engagement with the facility.
Deep gratitude is also extended to all individuals and institutions that have supported the operation, development, and community engagement of the Subaru Telescope. This research draws upon data related to the scholarly outputs and impacts of the Subaru Telescope, which is operated by the National Astronomical Observatory of Japan (NAOJ).
We are honored and grateful for the opportunity of observing the Universe from Maunakea, which has the cultural, historical, and natural significance in Hawaii.
Finally, the author dedicates this paper to his son Kimihiro and daughter Shiori, who organized a birthday celebration the day after the initial submission of this manuscript.
Parts of the manuscript text and code used for figure generation were assisted by ChatGPT (OpenAI). The authors are solely responsible for the final content and results.
\end{ack}



\appendix 

\section{FWCI and Outputs in Top Citation Percentiles}\label{app:indicators}

This appendix provides more detailed definitions and methodological notes for FWCI and Outputs in Top Citation Percentiles used in this study.

\subsection{FWCI}

\textbf{Definition (per publication):}  
The ratio of the actual number of citations received by a publication to the expected number of citations for publications of the same publication type, publication year, and subject field.

\textbf{Formula (for publication $P_i$):}  
\begin{eqnarray}
\mathrm{FWCI}_i = 
\frac{
  \text{Citation count of } P_i
}{
  \substack{
    \text{Average citation count of publications} \\
    \text{in the same year, field, and type as } P_i
  }
}
\end{eqnarray}

\textbf{Definition (for a publication set):}  
The FWCI for a group of publications is calculated as the average of the individual FWCI scores:
\begin{eqnarray}
\mathrm{FWCI}_{\text{set}} = \frac{1}{N} \sum_{i=1}^{N} \mathrm{FWCI}_{i}
\end{eqnarray}
where \( N \) is the number of publications in the set.

\textbf{Interpretation:}  
\begin{itemize}
  \item FWCI $= 1.0$: World average citation impact
  \item FWCI $> 1.0$: Above-average citation impact
  \item FWCI $< 1.0$: Below-average citation impact
\end{itemize}

\textbf{Notes:}  
FWCI is widely used in bibliometric evaluations because it normalizes for differences in citation behavior across disciplines, publication years, and publication types. This normalization enables meaningful comparisons across diverse research fields and time periods.

However, the version of FWCI implemented in SciVal is calculated based on citations received within a fixed three-year window following publication. For example, a paper published in 2005 is evaluated based on citations accumulated through the end of 2008. This time-limited window may fail to capture the full citation trajectory of certain papers, particularly in fields where citations accrue gradually over extended periods.

\subsection{Outputs in Top Citation Percentiles}

\textbf{Definition:}  
The number or proportion of publications that fall within the top $N$\% most cited papers, based on either raw citation counts or FWCI. In this study, we primarily use the FWCI-based metric for Outputs in the Top 10\% Citation Percentiles.

\textbf{Notes:}  
Like FWCI, FWCI-based Outputs in Top Citation Percentiles are calculated within a fixed citation window--three years after publication. As a result, both FWCI and the Top $N$\% citation percentiles may underrepresent the long-term impact of publications with delayed recognition. For more comprehensive evaluations, especially those focused on research infrastructure, it may be beneficial to incorporate complementary analyses using longer citation windows or field-specific benchmarks.

\end{document}